\documentclass{Interspeech}

\usepackage{cite}
\usepackage{amsmath,amssymb,amsfonts}
\usepackage{algorithmic}
\usepackage{graphicx}
\usepackage{textcomp}
\usepackage{xcolor}
\usepackage{booktabs}
\usepackage{url}
\usepackage{multirow}

\interspeechcameraready

\title{Pushing the Limits of End-to-End Diarization}

\author[]{Samuel J.}{Broughton}
\author[]{Lahiru}{Samarakoon}

\affiliation[nocounter]{}{Fano}{Hong Kong SAR, China}
\email{\{samuel.broughton,lahiru\}@fano.ai}

\newcommand{\vertical}[1]{\rotatebox[origin=c]{90}{#1}}
\usepackage{comment}

\begin{document}

\maketitle

\begin{abstract}
    In this paper, we present state-of-the-art diarization error rates (DERs) on multiple publicly available datasets, including AliMeeting-far, AliMeeting-near, AMI-Mix, AMI-SDM, DIHARD III, and MagicData RAMC. Leveraging EEND-TA, a single unified non-autoregressive model for end-to-end speaker diarization, we achieve new benchmark results, most notably a DER of 14.49\% on DIHARD III. Our approach scales pre-training through 8-speaker simulation mixtures, ensuring each generated speaker mixture configuration is sufficiently represented. These experiments highlight that EEND-based architectures possess a greater capacity for learning than previously explored, surpassing many existing diarization solutions while maintaining efficient speeds during inference.
\end{abstract}

\section{Introduction}

The perfect speaker diarization system should partition any given input audio signal into distinct segments, each correctly labeled with the corresponding speaker identity. In practice, this task is notoriously difficult due to the variability of input recordings. A robust system must handle varying degrees of background noise, a wide dynamic range, multiple speakers with diverse accents and languages, overlapping speech, and a broad range of audio quality captured from many different recording devices and conditions.

Traditional diarization systems are typically comprised of several components that are trained separately on non-diarization objectives \cite{park22dia_review}. Such cascaded pipelines usually include voice activity detection, speaker embedding extraction and clustering modules. Here, speaker embeddings are extracted from voice active frames so that a clustering algorithm can group together segments belonging to the same speaker. Generally, these solutions cannot handle overlapped regions of speech their reliance on multiple components increases both complexity and inference time.

End-to-end neural approaches to diarization directly address overlapping speech. End-to-end Neural Diarization (EEND) formulated diarization as a multi-label classification problem, outputting speaker labels for a fixed number of speakers \cite{fujita19eend, fujita19EEND-SA}. Many studies since enhanced EEND by adding a an Encoder Decoder Attractor (EDA) based calculation module to handle a variable number of speakers \cite{horiguchi20eend-eda, horiguchi22eend-eda}, and others have improved EDA with attention-based attractor mechanisms \cite{rybicka22end, fujita23intermediate, samarakoon23EEND-TA, chen2024attention, landini23diaper, harkonen2024eend}.

However, end-to-end neural networks such as EEND require a substantial amount of data to learn from, and annotated diarization datasets are scarce and expensive to label. To address this, is has become standard practice to generate simulated training mixtures with multiple speakers and pre-train models before fine-tuning on a real-world  dataset. Some works have further investigated the original simulation strategy proposed to construct the audio mixtures, developing an algorithm with the goal of better emulating aspects of natural conversations \cite{yamashita2022improving, landini2022simulated, landini2023multi}. This technique has been shown to increase diarization performance on telephony style data with little improvement on wide-band data \cite{chen2024attention}.

To date, few studies have examined the effects of simulating mixtures with a high number of speakers. One study used 2,500 hours of simulated audio with up to 10 speakers \cite{landini23diaper}, while another pre-trained on 16,700 hours with up to 18 speakers \cite{pan2024late}. However, the impact of simulating a higher number of speakers remains largely unexplored. 

In this work, we combine recent EEND-based advances with scaled pre-training, simulating mixtures for up to 8 speakers and increasing the total pre-training set to over 80,000 hours. This approach achieves state-of-the-art Diarization Error Rate (DER) for both end-to-end and cascaded systems across multiple datasets: AliMeeting-far (11.41\%), AliMeeting-near (8.55\%), AMI-Mix (11.04\%), AMI-SDM (15.16\%), DIHARD III (14.49\%) and MagicData RAMC (10.43\%).
	
\section{Related Work}

EEND-based methods have become the foundation of many diarization solutions. A common approach is to use an EEND module on short sliding windows to obtain local diarization results, then extract speaker embeddings and apply a clustering algorithm for global segmentation \cite{kinoshita21eend-vc, bredin2021end, bredin23pyannote21}.

To better address end-to-end use cases, EEND was extended with an EDA network, which outperforms conventional EEND and, in theory, handles an unlimited number of speakers. However, diarization performance degrades when the model encounters more speakers than seen during training \cite{horiguchi20eend-eda, horiguchi22eend-eda}.

Subsequent works have improved both the backbone and prediction head of EEND-EDA. Performance gains were demonstrated by replacing Transformers with Conformer layers \cite{liu21conformer, leung21conformer}, introducing Transformer-based attractor calculations \cite{rybicka22end, fujita23intermediate, samarakoon23EEND-TA, chen2024attention, landini23diaper, harkonen2024eend}, and using deep supervision training objectives \cite{fujita23intermediate, harkonen2024eend, landini23diaper}. Other approaches introduced supplementary inputs for further refinement, such as summary representations \cite{broughton23summary}, or diarization outputs for attractor estimation \cite{rybicka22end}.

Because end-to-end models require large amounts of training data, it is common practice to pre-train models on simulated mixtures \cite{fujita19eend, landini2022simulated, landini2023multi}. Increasing the number of simulated speakers within an audio mixture during pre-training can boost performance at fine-tuning \cite{horiguchi2021hitachi, chen2024attention}. 

\section{Method} \label{sec:method}

\subsection{Model Architecture: EEND-TA}

For this work, we selected EEND-TA as the network model architecture, a single unified non-autoregressive diarization model. We chose EEND-TA for its efficient design, which omits additional techniques such as iterative refinement \cite{rybicka22end}, or self-conditioning \cite{fujita23intermediate}. Previously, EEND-TA had only ever been shown to handle diarization for up to 4 speakers. The model consists of a Conformer encoder \cite{gulati2020conformer}, a combiner, and a Transformer decoder. Consistent with the original work, we use a modified Conformer encoder to accommodate the Conversational Summary Vector (CSV) \cite{broughton23summary}. This is a learnable special token that is prepended to the feature sequence before passing through the Conformer encoder. CSV was shown to improve performance on recordings with a higher number of speakers. The combiner joins the CSV with a set of learnable global embeddings, which are then input to the Transformer decoder. We use a standard Transformer decoder without positional embeddings to generate candidate speaker-wise attractors. Speaker existence posterior probabilities are then computed using a linear layer followed by a Sigmoid function. Final diarization outputs are computed by a Sigmoid function acting on the matrix product of the encoder output embeddings and speaker-wise attractors.

\subsection{Scaled Pre-Training}
\label{sec:pre-training}

We scaled pre-training by increasing the maximum possible number of speakers within a simulated mixture to 8. Simulated mixtures are generated by using a modified version of the standard simulated mixture algorithm \cite[Alg.~1]{fujita19eend}. 

Appropriate average silence interval values must be selected for each set of simulated mixtures, ranging from 1 to 8 speakers. Consistent with many other studies, for mixtures containing 1, 2, 3 or 4 speakers we applied the standard average silence interval values of $\beta = $ 2, 2, 5, 9, respectively. To determine $\beta$ values for mixtures containing 5, 6, 7 or 8 speakers, we calculated the average silence duration for each speaker in our real diarization dataset. We then took the mean silence duration for all speakers within each dataset split, where each dataset split represents recordings containing 5, 6, 7 or 8 speakers. This resulted in $\beta = $ 34, 54, 47, 50 for mixtures with 5, 6, 7 or 8 speakers, respectively.

However, using such large $\beta$ values can lead to simulated mixtures with extended periods of silence where no speaker is active. To address this, we modified the original algorithm to simply replace all silences longer than 5 seconds with a randomly chosen silence duration between 1 and 5 seconds.
 
\section{Experiments}

\begin{table*}[!h]
    \centering
    \caption{Diarization Error Rate (DER) across several datasets. Lower is better. Updated state-of-the-art results are marked in bold. Values in parenthesis are calculated with a 0.25 second forgiveness collar. ``+FT" denotes results for  fine-tuning on a single dataset.}
    \resizebox{\linewidth}{!}{
        \begin{tabular}{@{}llllllllll@{}}
            \toprule
            & \multicolumn{9}{c}{Dataset} \\ \midrule
            Model & \vertical{AISHELL-4} & \vertical{AliMeeting-far} & \vertical{AliMeeting-near} & \vertical{AMI-Mix} & \vertical{AMI-SDM} & \vertical{CALLHOME} & \vertical{DIHARD III} & \vertical{MagicData RAMC} & \vertical{VoxConverse} \\ \midrule
            VAD+VBx+OSD \cite{landini23diaper} & 15.84 & 28.84 & 22.59 & 22.42 & 34.61 & 26.18 & 20.28 & 18.19 & (6.12) \\
            EEND \cite{fujita19EEND-SA, horiguchi22eend-eda} & --- & --- & --- & 27.70 & --- & (21.19) & 22.64 & --- & --- \\
            EEND-EDA \cite{horiguchi22eend-eda} & --- & --- & --- & 15.80 & --- & (12.88) & 20.69 & --- & --- \\
            EEND-EDA\textsuperscript{§} (ours) & 13.43 & 12.30 & 9.05 & 11.31 & 16.29 & 17.62 & 15.02 & 10.60 & 15.98 \\
            DiaPer \cite{landini23diaper} & 29.0 & 20.7 & 17.8 & 23.9 & 40.7 & 24.16 & 21.1 & 16.1 & (19.1) \\
            pyannote.audio v3.1\textsuperscript{\textdagger} \cite{plaquet23pyannote31} & 12.2 & 24.4** & --- & 18.8 & 22.4 & 28.4 & 21.7 & 22.2 & 11.3 \\
            pyannoteAI\textsuperscript{\textdagger} & 11.9 & 22.5** & --- & 16.6 & 20.9 & 22.2 & 17.2 & 18.4 & 9.4 \\
            EEND-M2F \cite{harkonen2024eend} & 15.56 & 13.20 & 10.77 & 13.86 & 19.83 & 21.28 & 16.28 & 11.13 & 15.99 \\
            \quad +0.25s collar & (10.75) & (5.87) & (5.20) & (9.16) & (14.29) & (14.87) & (8.93) & (6.52) & (12.02) \\
            EEND-M2F + FT \cite{harkonen2024eend} & 13.98 & 13.40 & 10.45 & 12.62 & 18.85 & 23.44 & 16.07 & 11.09 & 16.28 \\
            \quad +0.25s collar & (9.34) & (6.11) & (5.02) & (7.92) & (13.33) & (16.72) & (8.82) & (6.46) & (12.36) \\ \midrule
            EEND-TA C4 (400 k) S4 & 25.68 & 12.68 & 10.36 & 12.88 & 20.08 & 19.24 & 17.50 & 10.97 & 21.75 \\
            EEND-TA C4 (400 k) S8 & 22.64 & 13.23 & 9.92 & 15.27 & 19.16 & 18.91 & 15.71 & 10.43 & 17.06 \\
            EEND-TA C4 & 15.41 & 11.73 & 9.29 & 14.56 & 17.68 & 19.21 & 15.13 & \textbf{10.30} & 17.03 \\
            EEND-TA\textsuperscript{§} & 15.09 & 11.45 & 9.05 & 11.31 & 16.13 & 17.51 & 15.29 & 10.57 & 15.75 \\
            EEND-TA & 15.31 & 12.65 & 8.60 & 11.06 & \textbf{15.16} & 16.91 & 14.76 & 10.43 & 15.44 \\
            \quad +0.25s collar & (10.82) & (6.83) & (4.09) & (7.19) & (10.19) & (10.97) & (8.16) & (5.79) & (11.57) \\
            EEND-TA + FT & 12.21 & \textbf{11.41} & \textbf{8.55} & \textbf{11.04} & 15.33 & 17.24 & \textbf{14.49} & 10.55 & 14.29 \\
            \quad +0.25s collar & (7.54) & (5.16) & (4.07) & (7.15) & (10.27) & (11.09) & (8.11) & (5.96) & (10.41) \\ \midrule
            State-of-the-art (as of Feb. 2025) & 11.7 & 13.20 & 10.45 & 12.62 & 15.4 & (\textbf{10.08}) & 16.07 & 11.09 & (\textbf{4.0})\textsuperscript{*} \\
            Source & \cite{han2025leveraging} & \cite{harkonen2024eend} & \cite{harkonen2024eend} & \cite{harkonen2024eend} & \cite{han2025leveraging} & \cite{chen23AED-EEND-EE} & \cite{he23ANSD-MA-MSE} & \cite{samarakoon23EEND-TA} & \cite{baroudi23pyannote_vox} \\ \bottomrule
        \end{tabular}
    }
    \vspace{-8pt}
    \begin{flushleft}
        \hspace{2pt} Some results from other works may use differing acoustic setups. \\
        \hspace{2pt}
        {}\textsuperscript{\textdagger} Results for pyannote systems are taken directly from \\ 
        \hspace{8pt} \url{https://github.com/pyannote/pyannote-audio/blob/develop/README.md}, as of commit 93ad8b9.\\
        \hspace{2pt}
        {}\textsuperscript{§} Only pre-trained EEND encoder weights loaded for fine-tuning, model head parameters are randomly initialized. \\
        \hspace{2pt}
        {}\textsuperscript{*} Potentially biased, as model was tuned and validated on VoxConverse test set.
    \end{flushleft}
    \label{res:main}
    \vspace{-5pt}
\end{table*}

\begin{table*}[!h]
    \centering
    \caption{Diarization Error Rate (DER) and Mean Speaker Counting Error (MSCE) per speaker on all datasets. Lower is better.}
    \resizebox{\linewidth}{!}{
        \begin{tabular}{@{}lllllllllllllllll@{}}
        \toprule
        Model & \multicolumn{2}{c}{1-spk} & \multicolumn{2}{c}{2-spk} & \multicolumn{2}{c}{3-spk} & \multicolumn{2}{c}{4-spk} & \multicolumn{2}{c}{5-spk} & \multicolumn{2}{c}{6-spk} & \multicolumn{2}{c}{7-spk} & \multicolumn{2}{c}{8-spk} \\ \midrule
        EEND-TA C4 (400 k) S4 & 7.68 & 0.20 & 9.36 & 0.03 & 16.37 & 0.45 & 17.02 & 0.44 & 30.55 & 1.43 & 23.02 & 2.23 & 24.15 & 3.28 & 27.31 & 4.19 \\
        EEND-TA C4 (400 k) S8 & 7.80 & 0.33 & 9.19 & 0.06 & 15.97 & 0.45 & 17.07 & 0.54 & 27.10 & 1.00 & 20.25 & 1.28 & 19.12 & 1.74 & 21.64 & 2.11 \\
        EEND-TA C4 & 8.21 & 0.47 & 9.60 & 0.06 & 15.07 & 0.40 & 14.93 & 0.45 & 21.27 & 0.86 & 16.46 & 0.97 & 16.62 & 1.52 & 23.55 & 1.30 \\ \midrule
        EEND-EDA\textsuperscript{§} (ours) & 7.47 & 0.23 & 9.01 & 0.04 & 14.72 & 0.37 & 14.71 & 0.42 & 20.06 & 0.89 & 16.69 & 1.10 & 15.84 & 1.61 & 20.01 & 2.22 \\
        EEND-TA\textsuperscript{§} & 7.14 & 0.23 & 8.97 & 0.04 & 14.35 & 0.41 & 14.55 & 0.38 & 21.56 & 0.89 & 16.78 & 1.03 & 16.27 & 1.65 & 20.55 & 2.07 \\
        EEND-TA & 8.86 & 0.27 & 9.02 & 0.07 & 15.02 & 0.40 & 14.02 & 0.34 & 23.81 & 1.00 & 14.55 & 0.97 & 13.47 & 1.22 & 18.96 & 1.56 \\ \bottomrule
        \end{tabular}
    }
    \vspace{-8pt}
    \begin{flushleft}
        \hspace{2pt}
        {}\textsuperscript{§} Only pre-trained EEND encoder weights loaded for fine-tuning, model head parameters are randomly initialized. \\
    \end{flushleft}
    \label{res:split}
    \vspace{-10pt}
\end{table*}

\begin{table}[!h]
\centering
    \caption{Train Dataset Statistics for 1 to 8 Speakers. ``Seen (\%)'' refers to the percentage of active speakers per utterance seen by the model across an entire epoch.}
    \resizebox{\linewidth}{!}{
        \begin{tabular}{@{}cllllll@{}}
        \toprule
        \multicolumn{1}{l}{} & \multicolumn{2}{c}{Overlap (\%)} & \multicolumn{2}{c}{Duration (hrs)} & \multicolumn{2}{c}{Seen (\%)} \\ \cmidrule(l){2-7} 
        \multicolumn{1}{l}{\# Spks} & \multicolumn{1}{c}{Real} & \multicolumn{1}{c}{Sim.} & \multicolumn{1}{c}{Real} & \multicolumn{1}{c}{Sim.} & \multicolumn{1}{c}{Real} & \multicolumn{1}{c}{Sim.} \\ \midrule
        1 & N/A & N/A & $2.7$ & $9047.1$ & $1.6$ & 12.5 \\
        2 & $5.0$ & $39.5$ & $246.2$ & $10,050.5$ & $41.4$ & 12.5 \\
        3 & $16.0$ & $38.0$ & $35.3$ & $12,757.3$ & $8.4$ & 12.5 \\
        4 & $25.0$ & $33.9$ & $321.1$ & $12,377.6$ & $36.3$ & 12.5 \\
        5 & $10.1$ & $13.7$ & $46.1$ & $10,130.5$ & $5.7$ & 12.5 \\
        6 & $8.2$ & $10.7$ & $4.1$ & $9821.4$ & $1.7$ & 12.5 \\
        7 & $9.0$ & $14.3$ & $17.0$ & $10,481.9$ & $2.5$ & 12.6 \\
        8 & $15.0$ & $15.6$ & $2.1$ & $11,083.9$ & $2.2$ & 12.4 \\ \bottomrule
        \end{tabular}
    }
    \label{table:dataset}
\end{table}
	
\begin{table}[!h]
    \centering
    \caption{Real Time Factor (RTF) was computed based on the total time to decode the DIHARD III evaluation set, one by one.}
    \resizebox{\linewidth}{!}{
        \begin{tabular}{@{}lccc@{}}
        \toprule
         &  & \multicolumn{2}{c}{RTF} \\ \cmidrule(l){3-4} 
        Model & \# Params (M) & CPU & GPU \\ \midrule
        pyannote.audio v3.1 \cite{plaquet23pyannote31} & $8.1^{\ddagger}$ & $3.5 \times 10^{-1}$ & $1.1 \times 10^{-2}$ \\
        EEND-EDA (ours) & $11.3$ & $2.7 \times 10^{-3}$ & $7.1 \times 10^{-4}$ \\
        EEND-M2F \cite{harkonen2024eend} & $16.3$ & $3.6 \times 10^{-3}$ & $2.5 \times 10^{-4}$ \\
        EEND-TA & $13.3$ & $2.2 \times 10^{-3}$ & $2.3 \times 10^{-4}$ \\ \bottomrule
        \end{tabular}
    }
    \vspace{-8pt}
    \begin{flushleft}
        \hspace{2pt}
        ${}^{\ddagger}$ Total parameters for the segmentation and embedding model
    \end{flushleft}
    \label{res:speed}
    \vspace{-15pt}
\end{table}

\subsection{Data}


Pre-training mixtures are generated from the LibriSpeech Corpus \cite{librispeech}, using the modified simulation strategy described in Section \ref{sec:pre-training}. Table \ref{table:dataset} compares the real datasets used at fine-tuning with the simulated mixtures used for pre-training. In total, we generated 100,000 simulated recordings for each possible number of active speakers in a mixture, resulting to 800,000 mixtures with 1 to 8 speakers. This amounted to over 80,000 hours of simulated data. There are two models shown in Tables \ref{res:main} and \ref{res:split} that use a total of 400,000 mixtures, ``EEND-TA C4 (400 k) S4 and ``EEND-TA C4 (400 k) S8''. The ``S4'' model is pre-trained with 1 to 4 speakers containing 100,000 mixtures per speaker, a setup more similar to the original work, whereas, the ``S8'' model is pre-trained with 1 to 8 speakers, using 50,000 simulated recordings per speaker mixture. 


For model fine-tuning, we use the publicly available datasets: AISHELL-4 \cite{aishell4}, AliMeeting \cite{alimeeting}, AMI-Mix (headset mixtures) and AMI-SDM (single distant microphone)  \cite{AMI}, CALLHOME \cite{callhome}, DIHARD III, MagicData RAMC \cite{ramc} and VoxConverse (version 0.3) \cite{voxconverse}. Where available, we use the official training, validation and testing splits provided by the dataset. For DIHARD III and VoxConverse we use the evaluation data for testing and split the train set into training and validation subsets using a 80\%:20\% split. The AMI-Mix and AliMeeting datasets containing multi-channel audio recordings are downmixed to a single channel. We refer to the AliMeeting far-field 8 microphone array recordings and headset microphone recordings as ``AliMeeting-far" and ``AliMeeting-near", respectively.
	
\subsection{Experimental Setup}

The input to EEND-TA is 23 log-dimensional Mel-filterbank features, extracted using a window length of 25ms and hop size of 10ms. The input feature sequence is first downsampled by a factor of 10 using convolutional layers with kernel sizes \{3, 5\}, strides \{2, 5\} and output feature dimension size 256. The downsampled sequence is then passed to a Conformer encoder consisting of 6 stacked layers each with 4 attention heads and feed-forward layers with 1024 hidden units. Before passing through the encoder, the input feature sequence is concatenated to a randomly initialized CSV representation. Models denoted as ``EEND-TA C4'' in Tables \ref{res:main} and \ref{res:split} use a Conformer encoder consisting of only 4 stacked layers. 

The Transformer Attractor calculation module consists of 3 Transformer decoder layers with 4 attention heads and a feed-forward dimension size of 1024. TA takes $S + 1$ randomly initialized 256 dimensional input queries, each of which are combined with CSV. Our models were trained to predict up to $S = 8$ speakers. Only the ``EEND-TA C4 (400 k) S4'' model was trained to predict up to $S = 4$ speakers.

The original EDA module was used to train ``EEND-EDA\textsuperscript{§} (ours)'' seen in Tables \ref{res:main} and \ref{res:split}. This model was trained to predict up to $S = 8$ speakers and also makes use of the CSV \cite{broughton23summary}.


Each Model was pre-trained for 2.5 M steps and fine-tuned for 250 k steps, all in bfloat16 precision. Pre-training used a batch size of 256 ($32 \times 8$) mixtures per step that were randomly cropped to an utterance length of 220 seconds. Fine-tuning used a batch size of 8 recordings cropped to 600 seconds. Chunk shuffling was applied only at fine-tuning \cite{leung21conformer}. The Adam optimizer and Noam scheduler with 100 k warm-up steps was used for pre-training, and the Adam optimizer with a fixed learning rate of $1 \times 10^{-5}$ was used for fine-tuning.

Model weights of the last pre-training checkpoint are used to initialize each model for fine-tuning. For ``EEND-EDA\textsuperscript{§} (ours)'' and ``EEND-TA\textsuperscript{§}'', we load only the pre-trained Conformer encoder from EEND-TA and randomly initialize the weights for the EDA and TA head modules before fine-tuning. The top 10 best checkpoints in terms of validation DER performance are averaged to create the model weights for inference. We conducted 20 k steps of dataset-specific fine-tuning after the aggregated fine-tuning stage of 250 k steps, marked as ``EEND + FT'' in Table \ref{res:main}. All models are evaluated using an attractor existence threshold of 0.5 and diarization threshold of 0.5.

\section{Discussion}

\subsection{Main Results}

The main results for our models, and current state-of-the-art results as of February 2025, are presented in Table \ref{res:main}. Unless stated otherwise, we report the least forgiving diarization error rates (DERs) by scoring all speech including overlaps, with no oracle speaker counting, no oracle voice activity detection, no hyper-parameter tuning on specific datasets and no forgiveness collar. It is important to note that each state-of-the-art result shown is obtained by fine-tuning a model on that specific dataset, except for AliMeeting-far, AMI-Mix and AMI-SDM.

Without dataset-specific fine-tuning, our model achieves state-of-the-art performance on 6 out of the 9 test sets: AliMeeting-far, AliMeeting-near, AMI-Mix, AMI-SDM, DIHARD III and MagicData RAMC. Impressively, EEND-TA improves the DER for DIHARD III by a relative 8.15\%. After dataset-specific fine-tuning (EEND-TA + FT) we further improve upon many DER results for a number of test sets.

There are three datasets where EEND-TA, with or without dataset-specific fine-tuning, does not beat current state-of-the-art results: AISHELL-4, CALLHOME and VoxConverse. Contrary to EEND-M2F, EEND-TA reaches near state-of-the-art performance on CALLHOME, while its results on AISHELL-4 and VoxConverse (containing 5 to 8 speakers and up to 21 speakers, respectively) are still comparable to EEND-M2F. Most of the errors made on these datasets stem from speaker confusion. After fine-tuning solely on AISHELL-4, our model achieves near state-of-the-art results.

\subsection{Scaling Pre-training}

By scaling model pre-training we show that EEND-TA had more capacity to learn than originally presented \cite{samarakoon23EEND-TA}. Table \ref{res:split} shows the progression of EEND-TA models before and after scaled pre-training per speaker. Looking at the ``S4'' and ``S8'' models, as expected, by including 5 to 8 speaker pre-training and allowing EEND-TA to predict up to 8 speakers, we see an improvement for both DER and MSCE for 5 to 8 speaker recordings with no major changes to results for 1 to 4 speakers. Doubling the pre-training data to 100,000 mixtures per speaker, ``EEND-TA C4'' further improves upon results, especially for recordings containing 3 to 7 speakers. Additionally, this model achieves the lowest DER on the MagicData RAMC test set.

An increase to 6 Conformer encoder layers also demonstrates a greater capacity to learn. Particularly for recordings containing 6 to 8 speakers, there are relatively large reductions to DER. Table \ref{res:split} also shows similar performance between EDA and TA variants. By comparing models that were both trained with randomly initialized heads, EEND-EDA shows consistent improvement for 5 to 8 speaker recordings. These results are competitive to models that fully load pre-trained model weights.

An important point to highlight is that even though our pre-training strategy crops each item in a batch to 220s, Table \ref{table:dataset} shows that for each epoch during pre-training the model sees sequences of 220s containing 8 speakers 12.4\% of the time.


Overall, the pre-training dataset has doubled in size, as has the maximum possible number of speakers within a single simulated mixture. This allowed for an increase to EEND-TAs total number of parameters by 30\% (10.2M to 13.3M). However, when compared to other speech domains, these models are still relatively small and lightweight. Given a much larger and more diverse simulated pre-training dataset, we hypothesize that diarization models can grow in size to match their ASR counterparts. Even though the pre-training dataset used here is approximately 80,000 hours in length, this can just be viewed an 80-fold augmentation of the original 960 hour long train set of the LibriSpeech Corpus. Therefore, future work will look towards further scaling the model size, pre-training dataset size, and pre-training dataset diversity.

\subsection{Real Time Factor}

When choosing a model for production, it is important to assess speed. Table \ref{res:speed} shows the Real Time Factor (RTF) when running inference on the DIHARD III evaluation set for best performing end-to-end diarization models and pyannote.audio v3.1. To calculate the RTF, we load the respective system and each recording of the 33 hour long evaluation set into memory. Audios are loaded as PyTorch Tensors. For the GPU benchmark, the diarization model or pipeline is pre-loaded into the GPU's vRAM. We then measure the total amount of time it takes each system to decode the entire entire list of Tensors, one by one. Both CPU and GPU benchmarks were conducted by using a single Intel Xeon Gold 6330 Processor. The GPU benchmark used a single NVIDIA RTX A6000.

As expected, end-to-end models are much faster at diarizing an audio when compared to clustering-based methods. EEND-TA is faster than EEND-M2F at both CPU and GPU decoding due to reduced model size and complexity. Here, EEND-TA processes the DIHARD III evaluation set around 460 times faster than real time with CPU decoding and 4290 times faster with GPU decoding, compared to pyannote.audio v3.1's more cascaded approach, which is only 3 and 92 times faster with CPU and GPU decoding,  respectively. EEND-EDA is slower than EEND-TA due to the sequential nature of LSTMs used in the EDA module. Given pre-processed log Mel-filterbank features, EEND-TA can process the entire combined 158 hours long test set in 97s on GPU, 5870 times faster than real time with GPU VRAM reaching a maximum of only 1.6 GiB.

\section{Conclusion}


In this paper, we demonstrated that end-to-end diarization models still possess a significant capacity for learning. By scaling model pre-training with EEND-TA, we achieved state-of-the-art results on AliMeeting-far, AliMeeting-near, AMI-Mix, AMI-SDM, DIHARD III, and MagicData RAMC. We also provided a strategy for generating up to 8-speaker simulation mixtures, so that a 220s random crop used for pre-training is highly likely to include all active speakers. Furthermore, our findings show that end-to-end methods yield substantially lower real-time factors, making them well suited to production environments. To the best of our knowledge, as of February 2025, the results presented in this paper outperform all other end-to-end diarization models.

\bibliographystyle{IEEEtran}
\bibliography{mybib}

\end{document}